\documentclass[11pt,twoside]{article}
\usepackage{asp2010}

\begin{document}

\title{Towards Non-spherical Radio Models}
\author{Val\'erio~A.~R.~M.~Ribeiro,$^1$ Wolfgang~Steffen,$^2$ Laura~Chomiuk,$^{3,4}$ \\ Nico~Koning,$^5$ Tim~J.~O'Brien,$^6$ and Patrick~A.~Woudt$^1$
\affil{$^1$Astrophysics, Cosmology and Gravity Centre, Department of Astronomy, University of Cape Town, Private Bag X3, Rondebosch, 7701, South Africa; vribeiro@ast.uct.ac.za}
\affil{$^2$Intituto de Astronom\'ia Universidad Nacional Aut\'onoma de M\'exico, C.P. 22860, Ensenada, Mexico}
\affil{$^3$Department of Physics and Astronomy, Michigan State University, East Lansing, MI 48824, USA}
\affil{$^4$National Radio Astronomy Observatory, P.O. Box O, Socorro, NM 87801, USA}
\affil{$^5$Department of Physics \& Astronomy, University of Calgary, Calgary, Alberta T2N 1N4, Canada}
\affil{$^6$Jodrell Bank Centre for Astrophysics, University of Manchester, Manchester M13 9PL, UK}}


\begin{abstract}
Radio observations of novae in outburst are of particular interest due to the physical parameters that may be retrieved from fitting the radio light curves. Most models that have fitted previous data assumed spherical symmetry however, it is becoming more and more clear that this is not the case. We explore morpho-kinematical techniques to retrieve the free-free radio light curves of non-spherical models and explore the effects of a non-spherical outburst on the physical parameters. In particular, we find that we may have been over estimating the ejected masses in the outburst of non-spherical novae.
\end{abstract}

\section{Introduction}
Nova ejecta have been resolved in the optical with a myriad of structures \citep*{H72,S83,SOD95,GO00,HO03,GO98,WSK09,R13}. For a few decades we have known that the outbursts are far from spherical. However, in the radio mostly spherically symmetric models of the outburst have been applied to observations. Most novae observed in the radio have been modelled as expanding, thermally emitting shells of ejecta, although some show evidence of non-thermal emission \citep{SB08}. The radio outburst is of particular interest due to the fact that we can derive properties of the ejecta since it is optically thick at much lower densities than other wavelengths and it does not suffer extinction. This provides us with a measure of the total ejected mass, density profiles, and kinetic energy \citep{H96}. Initial attempts to interpret the different observed structures was performed by \citet{H96} on V1974 Cygni (1992). These were aided with resolved radio images of the nova both during the optically thick and thin stages, which placed constraints on the kinematic behavior of the shells.

\section{Physics in {\sc shape}\footnote{Available from \url{http://bufadora.astrosen.unam.mx/shape/index.html}} }
In {\sc shape} \citep{SKW11} we construct an object interactively in a 3D interface. The structure is then transferred to a 3D grid on which radiation transfer is computed via ray tracing to the observer. Radiation transfer is based on emission and absorption coefficients which are provided as a function of physical parameters such as density, temperature and wavelength. As the rays emerge from the grid, images and spectra are generated. Temporal evolution is simulated when a model of the structure's expansion is provided. The time sequence of output is then generated automatically.
\begin{figure}[t]
\centering
\includegraphics[trim= 0 0 0 10, clip, scale=.4]{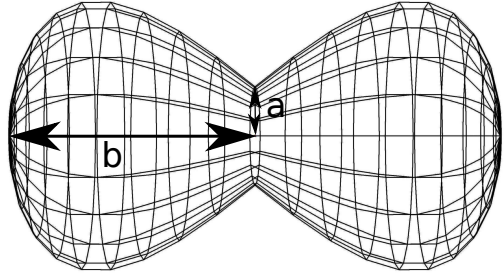}
\caption{Bipolar morphology (seen edge-on) as constructed in SHAPE. The semi-major (b) is four times the semi-minor (a) axis. The semi-major axes, expansion velocity and the mass were kept the same as in the spherical model.}
\label{fig1}
\end{figure}

We input in the physics module the free-free emission ($\epsilon_{\nu}$) and absorption coefficients ($\kappa_{\nu}$) at a given frequency ($\nu$), which are simplified to \citep{K07}:
\begin{displaymath}
4 \pi \epsilon_{\nu} = 6.8 \times 10^{-51} Z^2 N_e N_z T^{-0.5}_e \bar{g}_{ff}(\nu,T_e) \exp{-\frac{h\nu}{kT}} ,
\end{displaymath}
\begin{displaymath}
\kappa_{\nu} = 1.76 \times 10^{-12} Z^2 N_e N_z T^{-1.5}_e \nu^{-2}\bar{g}_{ff}(\nu,T_e).
\end{displaymath}
We take $N_e$ = $N_z$, the electron and ion densities, $Ze$ is the charge of the ion ($Z$ = 1 for a singly ionised atom) and $\bar{g}_{ff}(\nu,T_e)$ is the Gaunt factor:
\begin{displaymath}
\bar{g}_{ff}(\nu,T_e) = \left ( \frac{\sqrt{3}}{\pi} \right ) \left [ 17.7 + \ln \left ( \frac{T^{-1.5}_e}{\nu} \right ) \right ].
\end{displaymath}

We then modify the object to be bipolar with a ratio of the major to minor axis of four (Figure~\ref{fig1}). We explore different inclination angles, 0 degrees (the orbital plane is seen face on) and 90 degrees (the orbital plane is edge on). We assume a homologous expansion with a maximum velocity, $V_{\textrm{max}}$, of 414 km s$^{-1}$ electron temperature, $T_e$, of 17000 K, ejected mass, $M_{\textrm{ej}}$, of 1$\times$10$^{-4}$ M$_{\sun}$ and a 1/$r^2$ density distribution, on a spherical symmetric shell where the thickness was determined so that the inner radius was 0.25 times that of the outer radius. This is the same set up as \citet{HO07} on their models to replicate the 6-cm MERLIN observations of nova V723~Cas.

\section{Results}
\begin{figure}[t!]
\centering
\includegraphics[width=0.49\textwidth]{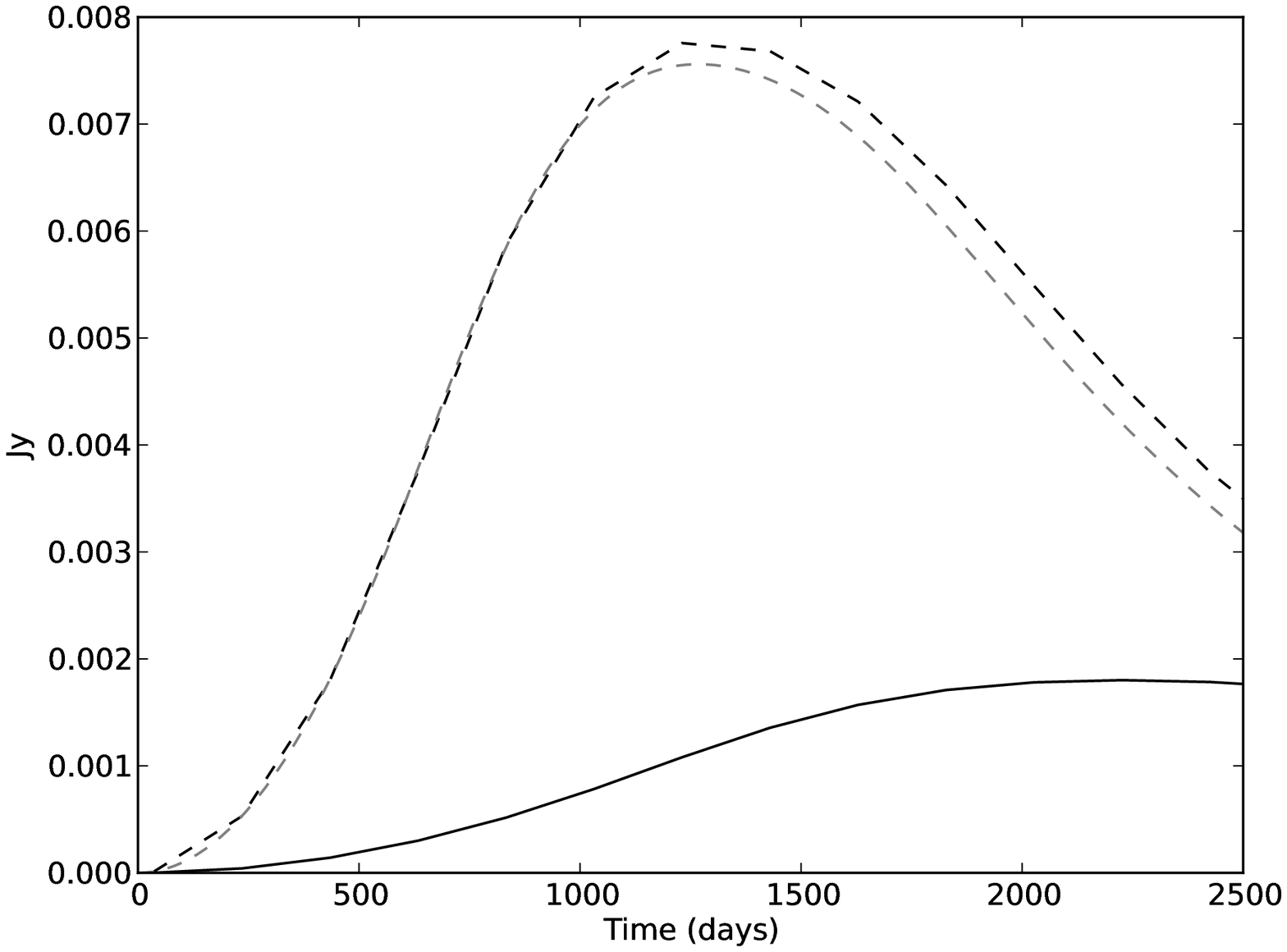}
\includegraphics[width=0.49\textwidth]{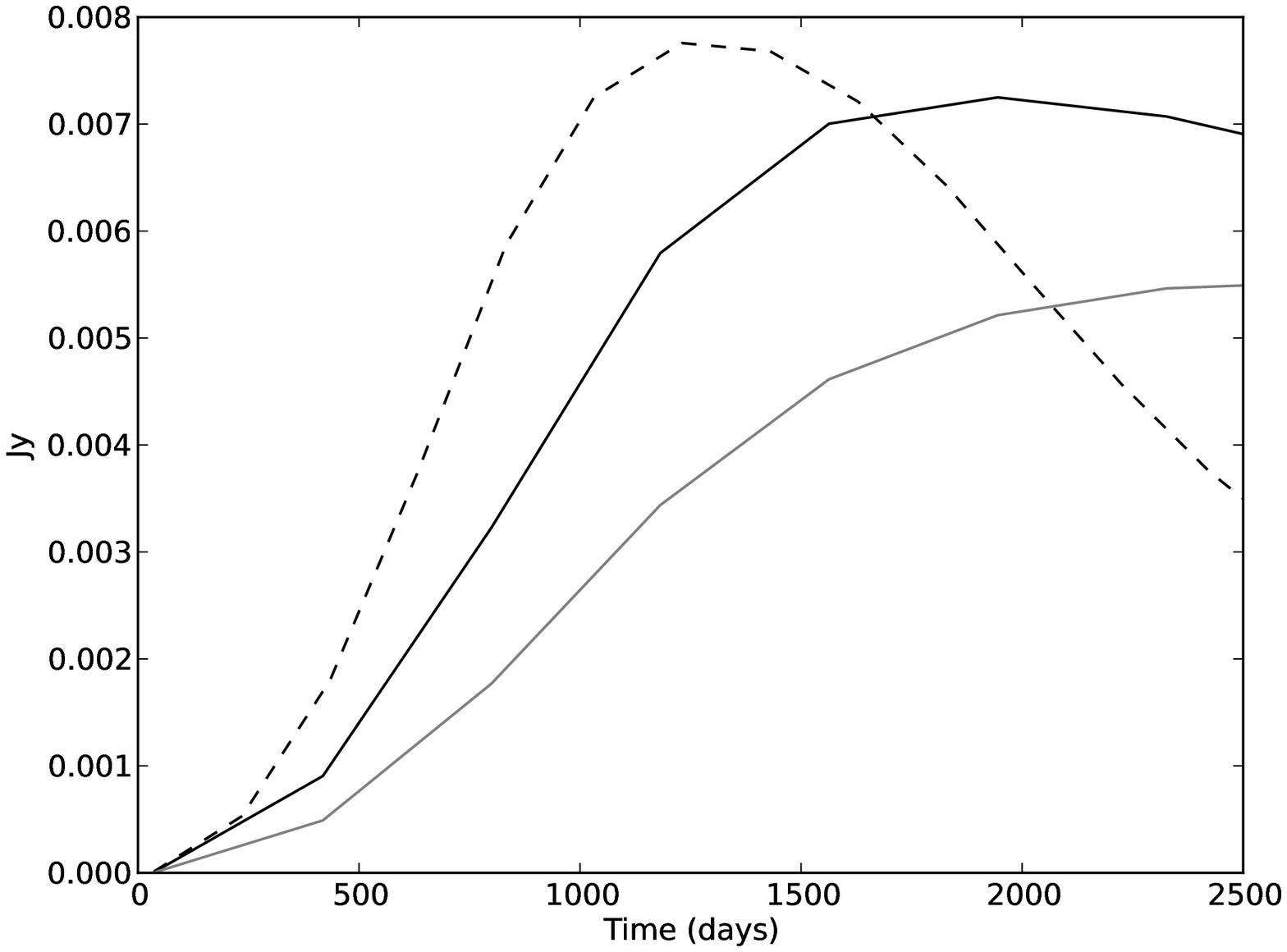}
\caption{Lightcurve models assuming $V_{\textrm{max}}$ = 414 km s$^{-1}$, $T_e$ = 17000 K, $M_{\textrm{ej}}$ = 1$\times$10$^{-4}$ M$_{\sun}$ and a 1/$r^2$ density distribution. \emph{Right} -- Dashed lines are spherical models presented for this work (black) compared with \citet{HO07} model (gray). The solid black line is for a similar spherical model however, in this case centered at 21-cm. \emph{Left} -- spherical model (dashed black) compared with bipolar models, assuming a ratio of the major to minor axis of 4, for inclination angles of 0 and 90 degrees (solid gray and black, respectively).}
\label{fig2}
\end{figure}

At first glance, Figure~\ref{fig2} (top) show our spherical model result with remarcable agreement with the model of \citet{HO07}; both in terms of the peak flux and turn over time. The slight disagreement exists due to how the mass is being estimated using {\sc shape}. Furthermore, we changed the central wavelength, to 21-cm (lower frequency) and as expected we have less flux and the lightcurve peaks later (solid black line).

In the bottom panel of Figure~\ref{fig2}, we show the spherical model (dashed line) comparing with a bipolar at 90 degrees (solid black) and 0 degrees (solid gray). During the initial optically thick rise phase, the models have the same steep rise, $t^{\alpha}$ (where $\alpha \sim$ 2), while the decline is much shallower for the bipolar models. The bipolar lightcurves also peak much later compared with the spherical model, while the bipolar at 0 degrees peaks even later at lower flux.

\section{Discussion}
The optically thick rise phase proceeds similarly in both the spherical and bipolar models. During the optically thin phase, in the spherical model, the ejecta becomes optically thin all throughout the sphere at the same time, yielding a steep decline; meanwhile in the bipolar case, material closer to the center of the explosion is much denser, and becomes optically thin later than the outer lobes. Furthermore, keeping all initial parameters the same, we find that the bipolar radio peak is much later than in the spherical case. These preliminary results demonstrates that we may have been over estimating the ejected masses in a few novae. Furthermore, we are at a time where optical emission line modelling, soon after outburst, are placing major constraints on the kinematic bahaviour of the shell \citep*[e.g.,][]{H96,RMV13}, which can then be used as input for these radio models \citep[see][for further details]{RCM14}.

\acknowledgements VARMR acknowledges the South African SKA Project for funding the postdoctoral fellowship at the University of Cape Town.


\end{document}